\documentclass[twocolumn,aps,showpacs,pra,10pt]{revtex4-1}

\usepackage{amsmath}
\usepackage{lipsum}
\usepackage{float}
\usepackage{txfonts}
\usepackage{graphicx}
\usepackage{mathptmx}
\usepackage{anyfontsize}
\usepackage{t1enc}
\usepackage{nicefrac}
\usepackage{color, hyperref}
\usepackage[squaren]{SIunits}

\newcommand{\be}{\begin{equation}}
\newcommand{\ee}{\end{equation}}
\newcommand{\ber}{\begin{eqnarray}}
\newcommand{\eer}{\end{eqnarray}}
\newcommand{\de}{\end{equation*}}
\newcommand{\cer}{\begin{eqnarray*}}
\newcommand{\der}{\end{eqnarray*}}

\begin{document}
\title{Divergence of optical vortex beams}
\author{Salla Gangi Reddy}
\affiliation{Physical Research Laboratory, Navarangpura, Ahmedabad, India-380 009.}
\email{sgreddy@prl.res.in}
\author{Chithrabhanu P}
\affiliation{Physical Research Laboratory, Navarangpura, Ahmedabad, India-380 009.}
\author{Shashi Prabhakar}
\affiliation{Physical Research Laboratory, Navarangpura, Ahmedabad, India-380 009.}
\author{Ali Anwar}
\affiliation{Physical Research Laboratory, Navarangpura, Ahmedabad, India-380 009.}
\author{J. Banerji}
\affiliation{Physical Research Laboratory, Navarangpura, Ahmedabad, India-380 009.}
\author{R. P. Singh}
\affiliation{Physical Research Laboratory, Navarangpura, Ahmedabad, India-380 009.}

\date{\today}

\begin{abstract}
 We show, both theoretically and experimentally,  that the propagation of optical vortices in free space can be analyzed by using the width ($w(z)$) of the host Gaussian beam and the inner and outer radii of the vortex beam at the source plane ($z=0$) as defined in \textit{Optics Letters \textbf{39,} 4364-4367 (2014)}. We also studied the divergence of vortex beams, considered as the rate of change of inner or outer radius with the propagation distance, and found that it varies with the order in the same way as that of the inner and outer radii at zero propagation distance. These results may be useful in designing optical fibers for orbital angular momentum modes that play a crucial role in quantum communication.
\end{abstract}

\maketitle

\section{Introduction}

Optical vortices or phase singular beams are well known due to their orbital angular momentum (OAM) \cite{allen, torner}. This OAM can be used as an information carrier which enhances the bandwidth as its modes have an infinite dimensional orthonormal basis \cite{padgett}. It has been shown that communication is possible with radio wave vortices \cite{thide1} and demonstrated experimentally for wireless communication \cite{thide2}. To use different OAM states for information processing, one should have both multiplexing and de-multiplexing setups for these modes and the same has been made with the use of fibers specially designed for OAM modes \cite{ram1, ram2, pravin}. For implementing these protocols with vortices, one should design fibers that support OAM modes in which case one should know the intensity distribution of these OAM modes as well as their divergence.

The spatial distribution of vortex beams has been studied under strong focussing conditions \cite{grier1, grier2} for which the position of maxima varies linearly with the order rather than square root of the order. The divergence of the vortex beams in free space propagation was studied theoretically by using the position of the radial variance of intensity \cite{andrews}. Recently, it has been shown theoretically that the divergence varies as the square root of the order if they are generated by using the pure mode converter \cite{astig} and it varies linearly \cite{padgett2} with the order if they are generated by using diffractive optical elements such as spatial light modulators \cite{cgh}. We have studied the spatial profile of vortices using two novel and measurable parameters (the inner and outer radii) and have given the analytical expressions \cite{gangi}. In this article, we study the divergence of vortices using these parameters experimentally as well as theoretically. We show that the inner and outer radii vary linearly with the propagation distance beyond the Rayleigh range. Furthermore, we show that the vortex at the plane of observation  depends simply on the width of the host Gaussian beam in that plane ($w(z)$) and on the inner and outer radii at $z=0$. To the best of our knowledge, this is the first experimental study on the divergence of optical vortex beams that is in close agreement with our theoretical predictions.

\section{Theory}

We start with the field distribution of a vortex of order $m$, embedded in a  Gaussian host beam of width $w_0$, as

\be
E_m(r)=(x+iy)^{m}\exp \left(-\frac{x^2+y^2}{w_0^{2}}\right)
\label{ov1}
\ee
and its intensity
\be
I_m(r)=r^{2m}\exp \left(-\frac{2r^2}{w_0^{2}}\right),\qquad r^2=x^2+y^2.
\label{ov}
\ee

\begin{figure}[h]
\begin{center}
\includegraphics[width=3.2in]{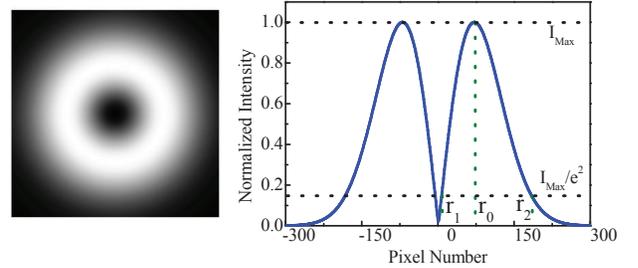}
\caption{ (Color Online) The intensity distribution (left) and its line profile (right) for an optical vortex beam of order 1.} \label{fig:fig1}
\end{center}
\end{figure}

This intensity distribution is shown in Fig. \ref{fig:fig1}. Here, we have defined two parameters for a vortex beam: inner and outer radii ($r_{1}, r_{2}$). These are the radial distances at which the intensity falls to $1/e^2 (13.6\%)$ of the maximum intensity at $r=r_0$ (say). Here, $r_{1}$ is the point closer to the origin or the center and $r_{2}$ is the point farther from the center, the outer region of the beam.
The distances $r_i$ ($i=0,1,2$) can be obtained as follows. For the sake of convenience, we set $w_0=1$, that is, $w_0$ is the unit of measuring radial distances. The inner and outer radii of the bright ring are given by \cite{gangi}
\begin{subequations}
\ber
r_1 (0)& = &(m+1.3-\sqrt{q_m})^{1/2}/\sqrt{2},\\
r_2 (0)& = &(m+1.3+\sqrt{q_m})^{1/2}/\sqrt{2},   \\
q_m &=&(m+1.3)^2-m^2\exp(-1.4/m) \nonumber.
\eer
\label{inner}
\end{subequations}

These radii correspond to the source plane ($z=0$) at which the vortices are being generated. The variation of inner and outer radii with propagation can be studied by propagating the vortex beam through free space. The field distribution of a vortex beam after propagating through an ABCD optical system is \cite{ashok}

\be
E_m(u,v)= \left(\frac{ikw^2_1}{2B}\right)^{m+1}(u+iv)^{m}\exp \left(-\frac{u^2+v^2}{w_2^{2}}\right) \\
\label{ovp}
\ee
where
\begin{subequations}
\ber
\frac{1}{w^2_1}& = & \frac{1}{w^2_0}+\frac{ikA}{2B},\\ \frac{1}{w^2_2} & = & \left(\frac{w_1k}{2B}\right)^2+\frac{ikD}{2B}.
\eer 
\label{i}
\end{subequations}

The corresponding intensity distribution is
\be
I_m(u,v)= \alpha_m r^{2m} \exp{(\frac{-2r^2}{\beta^2})} \\
\label{ovp1}
\ee
where
\be
\alpha_m= \frac{k^2w_1^2w_1^{*2}}{4B^2},\qquad r^2=u^2+v^2, \frac{2}{\beta^2}=\frac{1}{w_2^2}+\frac{1}{w_2^{*2}}.
\ee
 Here, $\alpha_m$ is a constant multiplicative factor and will not disturb the intensity pattern. Defining $\rho=r/\beta$, the above equation can be written as

\be
F_m = \frac{I_m}{\alpha_m\beta^{2m}} = \rho^{2m}\exp(-2\rho^2).\\
\label{ovp2}
\ee
This equation is similar to Eq. \ref{ov} and have the solutions $\rho_1$ and $\rho_2$ which are the same as that of $r_1(0)$ and $r_2(0)$ of Eq. \ref{inner}. Now the inner and outer radii of the vortex beam at a particular propagation distance are given by $\rho_1=r_1(z)=\beta r_1 (0)$ and $\rho_2=r_2(z)=\beta r_2 (0)$ where $r_1(0)$ and $r_2 (0)$ are the inner and outer radii at $z=0$. The expression for $\beta$ is
\be
\beta = w_0A\left(1+\frac{4B^2}{k^2A^2w_0^4}\right)^{1/2}. \\
\label{beta}
\ee
For free space propagation ($A=1$ and $B=z$), the above equation becomes

\be
\beta = w_0\left(1+\frac{4z^2}{k^2w_0^4}\right)^{1/2} = w_0\left(1+\frac{z^2}{z_R^2}\right)^{1/2}= w(z) \\
\label{beta}
\ee
where $z_R=\pi w_0^2/\lambda$ is the Rayleigh range and $w(z)$ is the width of the host Gaussian beam at distance $z$. Now, the inner and outer radii as a function of $z$ can be written as

\be
r_1(z)=w(z)r_1(0), \qquad r_2(z)=w(z)r_2(0).
\label{innz}
\ee

This relation shows that the defined parameters can describe the vortex beams as their propagation is similar to the host Gaussian beam.

Now, the divergence of optical vortices has been defined as the rate of change of inner and outer radii $r_{i}$, $i=1,2$ with respect to the propagation distance $z$ and is given by

\be
d_{m}^i (z) = \frac{\partial r_{i} (z)}{\partial z} = \frac{w_o r_i(0)}{\left(z^2+z_R^2)\right)^{1/2}}\frac{z}{z_R},\qquad i=1,2.
\label{d2}
\ee
 The divergence depends mainly on the corresponding radii at $z=0$. At large $z$ ($z>>z_R$), the divergence is constant and is given by

\be
d_m^i = \frac{w_or_i(0)}{z_R}.\\
\label{d3}
\ee

These results are same for vortex beams either they are generated by using mode converter or diffractive optical elements as we consider only the intensity distribution \cite{padgett2}.

\begin{figure}[h]
\begin{center}
\includegraphics[width=3.2in]{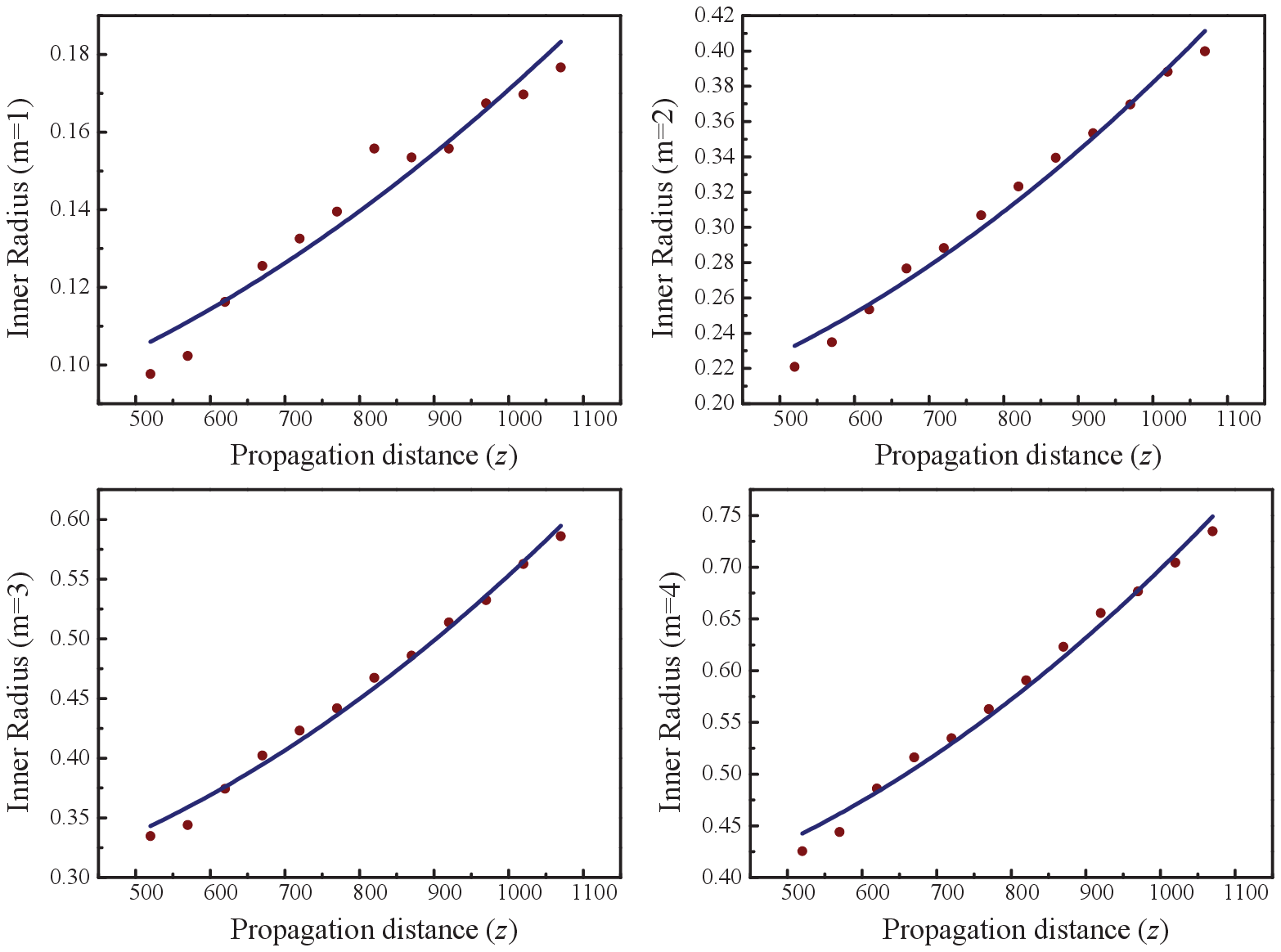}
\caption{ (Color Online) The inner radius for the optical vortex beams of orders 1--4 at different propagation distances. The experimental data (red dots) and the corresponding fitting (blue solid line) with Eq. \ref{innz}. } \label{fig:fig4}
\end{center}
\end{figure}

\begin{figure}[h]
\begin{center}
\includegraphics[width=3.2in]{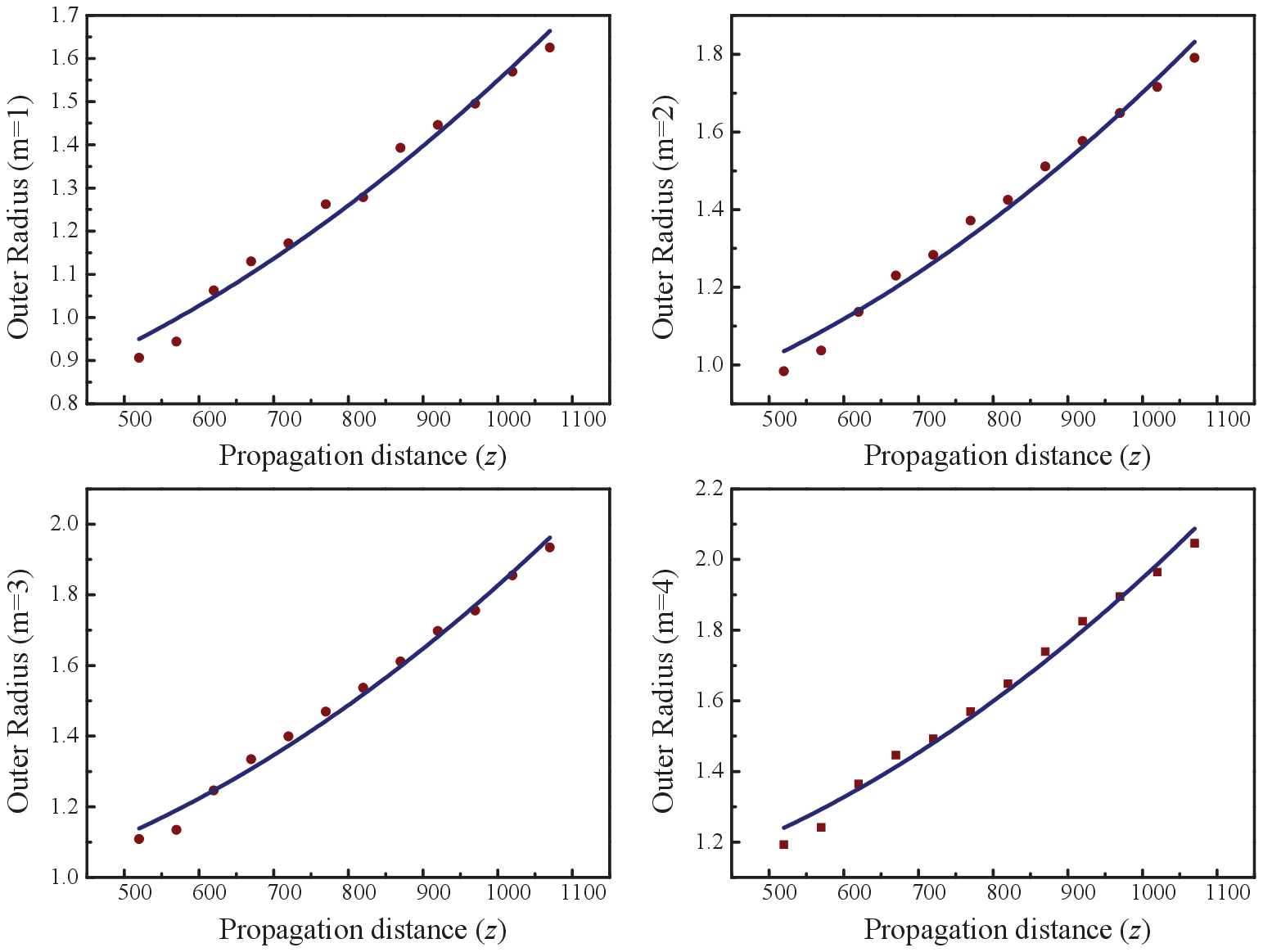}
\caption{ (Color Online) The outer radius of optical vortices with orders 1--4 at different propagation distances. The experimental data (red dots) and the corresponding fitting (blue solid line) with Eq. \ref{innz}.} \label{fig:fig5}
\end{center}
\end{figure}

\section{Results and discussion}

We use an intensity stabilised He-Ne laser to generate optical vortex beams using computer generated holography technique. We introduce holograms corresponding to the vortices of different orders to the spatial light modulator (SLM) (Holoeye LCR-2500). We then allow the Gaussian laser beam to be incident normally on the SLM at the branch points of hologram which gives the vortex beam in the first diffraction order \cite{gangi}. The vortices of different orders from 1 to 5 have been recorded at different propagation distances starting from 52 cm to 107 cm in steps of 5 cm using an Evolution VF color cooled CCD camera of pixel size 4.65 $\micro$m. We have recorded 20 images for a given order and at a given propagation distance. These images have been further processed to determine the inner and outer radii.

Figures \ref{fig:fig4} and \ref{fig:fig5} show the variation of inner and outer radii with propagation distance for optical vortex beams of orders $m$ = 1--4. Dots represent the experimental data and the solid lines represent the fitting with Eq. \ref{innz}. From the figures, it is clear that the experimental results are in excellent agreement with the theoretical predictions. The slope of these curves at $z>>z_R$ has been taken as the divergence i.e. the rate of change of inner and outer radii. The experimental results are obtained by taking the average over 20 images and the error bars are too small to be visible in the plot.

\begin{figure}[h]
\begin{center}
\includegraphics[width=2.6in]{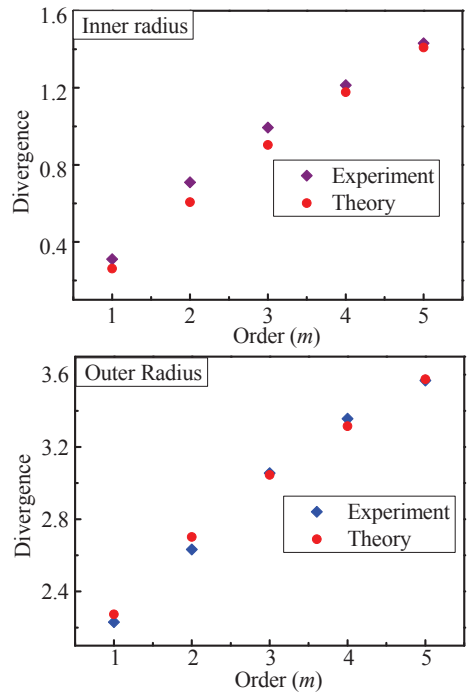}
\caption{ (Colour online) The divergence i.e. the rate of change of inner (top) and outer (bottom) radii with the order of vortex beams obtained experimentally (blue dots) as well as theoretically (red dots). } \label{fig:fig6}
\end{center}
\end{figure}

Figure \ref{fig:fig6} shows the variation of divergence with order of the vortex. The rate of change of inner and outer radii are directly proportional to their value at $z=0$. The blue diamonds represent experimentally obtained divergence and the red dots represent the corresponding results obtained from Eq. \ref{d3}. The variation of divergence with order for the inner and outer radii is the same as the variation of corresponding radius with order at $z=0$ as is clear from Eq. \ref{d3}.

\section{Conclusion}
In conclusion, we have described the intensity distribution of optical vortices by using two measurable parameters, the inner and outer radii. The propagation dynamics of vortices can be easily described by its intensity distribution at the source plane and the width of the host Gaussian beam at the plane of observation. We have also studied the rate of change of inner and outer radii, divergence, and their variation with the order. The experimental results are well supported by the analytical results. These results may be useful in designing fibers specially for the orbital angular momentum modes.

\end{document}